\documentstyle[prl,aps,epsf]{revtex}

\newcommand{\Szz}{{\rm S}_{(z \leftrightarrow -z)}}
\begin{document}
\draft
\title{Comments on 
`Existence of axially symmetric solutions in SU(2)-Yang Mills
and related theories [hep-th/9907222]'}
\author{
{\bf Burkhard Kleihaus}}
\address{NUIM, Department of Mathematical Physics, Maynooth, Co. Kildare,
Ireland\\
}
\author{and}
\author{
{\bf Jutta Kunz}}
\address{Fachbereich Physik, Universit\"at Oldenburg, D-26111 Oldenburg,
Germany
}
\date{\today}
\maketitle
\begin{abstract}
In [hep-th/9907222] Hannibal claims to exclude the existence 
of particle-like static axially symmetric non-abelian solutions 
in $SU(2)$ Einstein-Yang-Mills-dilaton theory.
His argument is based on the asymptotic behaviour of such solutions.
Here we disprove his claim by giving explicitly
the asymptotic form of non-abelian solutions with winding number $n=2$.
\\
\\
\end{abstract}
\bigskip
Particle-like
static axially symmetric solutions of Yang-Mills-dilaton (YMD) \cite{YMD}
and Einstein-Yang-Mills-dilaton (EYMD) theory \cite{EYMD1,EYMD2} have been 
investigated numerically and analytically \cite{regu} in recent years. 
The gauge potential, given in a singular gauge in
\cite{YMD,EYMD1,EYMD2}, can locally be gauge 
transformed into a regular form \cite{regu}. 
On intersecting neighbourhoods the regular gauge potentials 
can be gauge transformed into each other by regular gauge transformations. 
Hence these solutions are globally regular.

Recently, Hannibal \cite{han1}
claimed to have shown that the static axially symmetric solutions
discussed in \cite{YMD,EYMD1,EYMD2,EYMDBH1,EYMDBH2,regu} are singular
in the gauge field part.
However, in \cite{han1}
he only observed that the {\em gauge potential} of 
refs.~\cite{YMD,EYMD1,EYMD2,EYMDBH1,EYMDBH2} does not obey 
a set of {\em sufficient} regularity conditions \cite{RR}.
Repeating his claim in \cite{han2},
and noting that the ``singular'' solutions 
can be locally gauge transformed into regular solutions
as discussed in \cite{regu,rep1},
Hannibal apparently uses the equality 
$solution = gauge\  potential$, despite the fact that the 
gauge potential is not uniquely defined and the {\em same} solution
can be given in many {\em different} gauges.

In \cite{han2} Hannibal 
then turns to the question of existence
of static axially symmetric solutions of YMD and EYMD theory.
He presents as his results, that i) 
``there exist only embedded abelian particle-like solutions''
and ii)
``the solutions constructed by Kleihaus and Kunz are 
shown to be gauge equivalent to these''.

Particle-like solutions of EYMD  and related theories -- which are not 
embedded abelian solutions -- have been investigated in the past decade by 
several authors (see e.~g. \cite{others}) and it is generally accepted that 
these solutions are genuine non-abelian solutions. For rigorous proofs 
of the existence of these solutions see e.~g.~\cite{proofs}.

Concerning the static axially symmetric solutions, the argument of Hannibal 
relies on the fact that he was not able to find asymptotic solutions, which 
possess the correct symmetries with respect to the discrete transformation
\begin{equation}
\Szz : \theta \rightarrow \pi-\theta \ .
\label{S}
\end{equation}
However, he chooses a gauge in which this symmetry is lost for all non-trivial
gauge field functions parameterizing the gauge potential.
Apparently he takes this into account for one of the functions,
but does not take it into consideration for the other two functions.
Without taking into account the correct symmetry behaviour of the 
gauge field functions the results are not reliable. 

By giving a counterexample,
we here disprove Hannibal's claim,
{\sl that ``there exist only embedded abelian 
particle-like solutions ...'' \cite{han2}, because asymptotic solutions for 
genuine non-abelian gauge potentials do not exist}.
This counterexample represents
the correct asymptotic behaviour for a {\sl non-abelian}
static axially symmetric solution of EYMD theory
\cite{EYMD1,EYMD2,EYMDBH1,EYMDBH2}.
 
Using polar coordinates ($r$, $\theta$, $\varphi$),
we parameterize the static axially symmetric $su(2)$ gauge potential as
\cite{EYMDBH1}
\begin{equation}
A_\mu dx^\mu = \frac{1}{2g}\left\{
\left[\frac{H_1}{r} dr +(1-H_2)d\theta\right] 
\tau^n_\varphi
-n \sin\theta \left[ H_3\tau^n_r +(1-H_4)\tau^n_\theta \right] d\varphi
\right\} \ ,
\label{ansatz}
\end{equation}
where $H_i$ are functions of $r$ and $\theta$ and $n$ denotes the 
winding number. For convenience we 
set the gauge coupling constant $g$ equal to one in the following.
The $su(2)$ matrices $\tau^n_\varphi,\tau^n_r,\tau^n_\theta$ are 
defined in terms of Pauli matrices $\tau_1,\tau_2, \tau_3$ by
\begin{equation}
\tau^n_\varphi = -\sin(n\varphi) \tau_1+\cos(n\varphi) \tau_2 \ ,
\tau^n_r = \sin\theta \tau^n_\rho + \cos\theta \tau_3 \ ,
\tau^n_\theta = \cos\theta \tau^n_\rho - \sin\theta \tau_3 \ ,
\end{equation}
with $\tau^n_\rho= \cos(n\varphi) \tau_1+\sin(n\varphi) \tau_2$.

In order to compare with ref.~\cite{han2} we parameterize the functions $H_i$  
as \cite{han1}
\begin{eqnarray}
H_1 & = & \left[\tilde{F}_1 \sin^2\theta +\left(n f(r) +\tilde{F}_2\right)
\right]
\cos\theta \sin^{|n|}\theta  \ ,
\label{h1reg} \\
(1-H_2) & = & \left[ n f(r) 
+ \tilde{F}_1 \sin^2\theta\cos^2\theta 
-\left(n f(r)+\tilde{F}_2\right) \sin^2\theta\right]
\sin^{|n|-1}\theta  \ ,
\label{h2reg} \\
H_3 & = & 
\left[
\left( f(r) + \tilde{F}_3 \sin^2\theta\right)\cos\theta\sin^{|n|-1}\theta 
 \right] \sin\theta
+\tilde{F}_4  \sin\theta\cos\theta 
\label{h3reg} \\
 & = & F_3 \sin\theta + F_4 \cos\theta\ ,
\label{f3reg} \\
(1-H_4) & = & 
\left[
\left(f(r) + \tilde{F}_3 \sin^2\theta \right)\cos\theta\sin^{|n|-1}\theta 
 \right]\cos\theta
-\tilde{F}_4 \sin^2\theta 
\label{h4reg} \\
 & = & F_3 \cos\theta - F_4 \sin\theta\ ,
\label{f4reg} 
\end{eqnarray}
where $\tilde{F}_i$ are continuous functions of $r$ and $\theta$ and
$F_3 =( f(r) + \tilde{F}_3 \sin^2\theta)\cos\theta\sin^{|n|-1}\theta $, 
$F_4 =\tilde{F}_4 \sin\theta$ have been introduced for later convenience.
Assuming that $\tilde{F}_i$ are regular functions of $r^2$ and 
$\sin^2\theta$ \cite{RR}, this parameterization of the ansatz guarantees 
that the gauge potential is regular on the $z$-axis ($|z|>0$) and that 
the functions $H_1$ and $H_3$ are odd under the transformation 
$\Szz : \theta \rightarrow \pi-\theta$, whereas the 
functions $H_2$ and $H_4$ are even \cite{han1}.
In order to guarantee regularity at the origin, additional conditions have 
to be imposed on the functions  $\tilde{F}_i$. 

The gauge potential (\ref{ansatz}) is form invariant under abelian gauge 
transformations of the form \cite{regu}
\begin{equation}
U = \exp{\left\{i \Gamma \tau^n_\varphi/2\right\}} \ ,
\label{gam}
\end{equation}
where $\Gamma$ is a function of $r$ and $\theta$.
The functions $H_i$ transform like
\begin{eqnarray}
{H_1} &\longrightarrow &\hat{H}_1 
= H_1 - r \partial_r \Gamma  \ ,
              \label{ET_H1}\\
{H_2} &\longrightarrow &\hat{H}_2 
= H_2 +  \partial_\theta \Gamma  \ ,
              \label{ET_H2}\\
{H_3} &\longrightarrow & \hat{H}_3 
= \cos \Gamma (H_3+\cot \theta) -\sin \Gamma H_4 - \cot \theta 
             \ ,   \label{ET_H3}\\
{H_4} &\longrightarrow & \hat{H}_4 
= \sin \Gamma (H_3+\cot \theta) +\cos \Gamma H_4 
             \ .  \label{ET_H4}\\
\nonumber
\end{eqnarray}
Following \cite{han2},
we now fix the gauge such that $\hat{H}_2=1$ and assume that 
$H_2$ is given in the form (\ref{h2reg}). The gauge transformation
function is given by 
\begin{equation}
\Gamma(r,\theta) = 
\int_{\theta_0}^\theta \left\{1-H_2(r,\theta')\right\}d\theta' \ .
\label{gamh2}
\end{equation}
Again following \cite{han2},
we choose $\theta_0 = 0$, 
in order to maintain regularity on the positive $z$-axis ($z > 0$).
Then the gauge potential may diverge on 
the negative $z$-axis \cite{han2}.
Along the (positive) $z$-axis
the function $\hat{H}_1$ is still of the order $O(\sin^{|n|}\theta)$ 
and may be written locally as 
$\hat{H}_1=\bar{F}_1\cos\theta \sin^{|n|}\theta$,
with some function $\bar{F}_1(r,\theta)$. 
The gauge transformed function $\hat{F}_3$ is now of order 
$O(\sin^{|n|+1}\theta)$ and may be written as
$\hat{F}_3=\bar{F}_3 \cos\theta\sin^{|n|+1}\theta$ near the (positive)
$z$-axes, with some function $\bar{F}_3(r,\theta)$.
Hence, we find near the (positive) $z$-axis
\begin{eqnarray}
\hat{H}_1 & = & \bar{F}_1 \cos\theta \sin^{|n|}\theta  \ ,
\label{h1_h2} \\
(1-\hat{H}_2) & = & 0 \ ,
\label{h2_h2} \\
\hat{H}_3 & = & 
\left[\bar{F}_3 \sin^{|n|+1}\theta
 +\bar{F}_4 \right] \sin\theta\cos\theta\ ,
\label{h3_h2} \\
(1-\hat{H}_4) & = & 
\bar{F}_3 \cos^2\theta\sin^{|n|+1}\theta 
-\bar{F}_4 \sin^2\theta \ .
\label{h4_h2}
\end{eqnarray}
This form of the gauge field functions $\hat{H}_i$ could have been obtained from
Eqs.~(\ref{h1reg})-(\ref{h4reg}) by setting $f(r)\equiv 0$ and
$\tilde{F}_2=\cos^2\theta \tilde{F}_1$ \cite{han2}. 
However, under the assumption 
that the functions $\tilde{F}_i$ are continuous everywhere,
this would not have been correct.
This can be seen as follows.

The function $\Gamma$ contains an even part and an odd part with
respect to the transformation $\Szz$ \cite{han2}. 
As a consequence the 
function $\hat{H}_1$ is no longer odd with respect to $\Szz$ \cite{han2} and 
cannot be written in the form (\ref{h1_h2}) with continuous functions 
$\bar{F}_i$. It is evident from Eqs.~(\ref{ET_H3})-(\ref{ET_H4}) that the 
functions $\hat{H}_3$ and $\hat{H}_4$ also have lost their antisymmetry, 
respectively symmetry,  with respect to $\Szz$.
Thus all the functions $\hat{H}_i$ may take finite 
values on the $\rho$-axis. 
If we then insist to use the parameterization 
Eqs.~(\ref{h1_h2}-\ref{h4_h2}) for the gauge transformed functions $\hat{H}_i$,
not only near the (positive) $z$-axis but everywhere, 
we must allow the functions $\bar{F}_1$ and $\bar{F}_3$ to contain 
the factor $1/\cos\theta$.

Parameterizing the metric \cite{EYMD1,EYMD2,EYMDBH1,EYMDBH2,han2} by 
\begin{equation}
ds^2= -{\rm g}(r,\theta) dt^2
      +\frac{{\rm m}(r,\theta)}{{\rm g}(r,\theta)}(dr^2 + r^2 d\theta^2)
      +\frac{{\rm l}(r,\theta)}{{\rm g}(r,\theta)} r^2 \sin^2\theta d\varphi^2
      \ ,
\label{metric}            
\end{equation}
we substitute the gauge transformed ansatz Eq.~(\ref{ansatz}) 
together with (\ref{h1_h2})-(\ref{h4_h2}) and (\ref{metric}) into the 
coupled Einstein, dilaton and Yang-Mills equations and
find for the asymptotic EYMD solution with winding number $n=2$,
\begin{eqnarray}
\bar{F}_1 & = & 
-\frac{{\rm C}_2}{r^2} \frac{1-\cos\theta}{\cos\theta\sin^2\theta} + (\dots) \ ,
\label{tf1} \\
\bar{F}_3 & = & 
\frac{{\rm C}_2}{r^2}
\frac{2(1-\cos\theta)-\cos\theta\sin^2\theta}{4\cos\theta\sin^4\theta} 
       + (\dots) \ ,
\label{tf3} \\
\bar{F}_4 & = & 
\frac{{\rm C}_1}{r}+\frac{{\rm C}_1}{4 r^2}(\bar{{\rm g}}_1-2\phi_1)+(\dots)\ ,
\label{tf4} 
\end{eqnarray}
for the gauge field functions and
\begin{equation}
\Phi  =  \frac{\phi_1}{r}+ (\dots)\ , \ \ \ \
{\rm g}  =  1 + \frac{\bar{{\rm g}}_1}{r} +\frac{\bar{{\rm g}}^2_1}{2r^2}
+(\dots)\ , \ \ \ \
{\rm l}  =  1 + \frac{\bar{{\rm l}}_2}{r^2}
+(\dots)\ , \ \ \ \
{\rm m}  =  1 + \frac{\bar{{\rm l}}_2+\bar{{\rm m}}_2\sin^2\theta}{r^2}
+(\dots)\ ,
\label{tfphi} 
\end{equation}
for the dilaton function \cite{dil} and the metric functions,
where
${\rm C}_1$, ${\rm C}_2$, $\phi_1$, 
$\bar{{\rm g}}_1$, $\bar{{\rm l}}_2$, $\bar{{\rm m}}_2$
are constants \cite{related}.  The $(\dots)$ 
indicate terms vanishing faster than $1/r^2$, which can not necessarily be 
expanded in powers of $1/r$. It is easy to see that $\bar{F}_1$ and
$\bar{F}_3$ are finite on the positive $z$-axis and can be expressed 
as polynomials in $\sin^2\theta$ in the vicinity of the $z$-axis.
Hence, the asymptotic solution is regular on the positive $z$-axis. 
On the $\rho$-axis the gauge field functions $\hat{H}_i$ are finite, 
as expected. 
Furthermore, $\hat{H}_1$ can be decomposed into an odd part and an even part 
with respect $\Szz$, where the latter is a function of $r$ only \cite{han2}. 
As a consequence all derivatives 
${\displaystyle \frac{\partial^{2k}\hat{H}_1}{\partial \theta^{2k}}}$ vanish on the 
$\rho$-axis.
On the $z$-axis the functions $f_{10}(r):=\bar{F}_1(r,\theta=0)$ and 
$f_{40}(r):=\bar{F}_4(r,\theta=0)$ are not necessarily zero.
Hence, this solution is not an embedded abelian solution and has to be 
classified as type III according to ref.~\cite{han2}. 
For winding number $n=3,4$ we have
constructed asymptotic solutions, too, which 
possess all the correct symmetry and regularity properties.
These simple solutions disprove the claim of Hannibal \cite{han2} that only 
for embedded abelian gauge potentials asymptotic solutions with the correct
symmetries can exist in EYMD and related theories.

We have also analysed the solution with winding number $n=2$ in the 
Coulomb gauge $r H_{1,r}-H_{2,\theta}=0$ \cite{asymp}, which was 
employed in the numerical construction of the solutions 
\cite{YMD,EYMD1,EYMD2,EYMDBH1,EYMDBH2}. In this gauge the gauge 
potential is not well defined on the $z$-axis. However,
transforming the asymptotic solution obtained in the Coulomb gauge into
the gauge $H_2=1$, we find the same form  Eqs.~(\ref{tf1}-\ref{tf4}).
This shows that at infinity the gauge potential in the Coulomb gauge can be 
locally gauge transformed into a regular gauge potential and that the 
asymptotic gauge potentials in the different gauges correspond to the 
same asymptotic solution. 
Note, that in the Coulomb gauge
the functions $H_1$ and $H_3$ are antisymmetric, whereas the functions
$H_2$ and $H_4$ are symmetric with respect to $\Szz$. 

In Ref.~\cite{han2}
Hannibal failed to obtain asymptotic non-abelian solutions such as
the solution Eqs.~(\ref{tf1})-(\ref{tfphi}).
Clearly, he rejected solutions because the functions $\hat{H}_i$ 
did not possess the symmetry property he erroneously expected.
Furthermore, he tried to find the solutions in form of power series in
$\sin\theta$. However, the solution Eqs.~(\ref{tf1})-(\ref{tf3})
contains the function $1/\cos\theta$, which, 
considered as a power series in $\sin\theta$, has radius of convergence 
$|\sin\theta|<1$. Thus, the results at $\theta=\pi/2$ may be not 
reliable. Unfortunately, the presentation in \cite{han2}
is not consistent at various places,
and it remains unclear which calculations exactly were
carried out and whether additional sources of error
appeared in the analysis.

We close with some
additional comments on ref.~\cite{han2}:

It is assumed in \cite{han2} that the 
asymptotic solutions can be obtained as a power 
series in $1/r$. However, if one wants to exclude the existence of all 
solutions (except embedded abelian solutions),
then a proof that all solutions are 
analytic in the variable $1/r$ at $1/r=0$ would be necessary. 
Such a proof is not 
given in ref.~\cite{han2}. Indeed, we find that in the Coulomb gauge
terms in higher order arise, which can not be expanded in a power series in 
$1/r$ \cite{asymp}.

It is claimed in \cite{han2} that there exists
a regular, globally defined gauge potential for
the solutions of ref.~\cite{YMD,EYMD1,EYMD2,EYMDBH1,EYMDBH2}.
However, the ``proof'' of existence of such a gauge 
potential is not complete, because no attention is paid to the regularity of 
the gauge potential at the origin \cite{regu}. It can be seen easily, that 
the gauge potential is not twice differentiable at the origin,
if high powers in $\sin\theta$ arise, which are not multiplied by 
sufficiently high powers in $r$.

In \cite{han2} the static axially symmetric solutions of 
ref.~\cite{YMD,EYMD1,EYMD2,EYMDBH1,EYMDBH2}
are classified as gauge equivalent to 
type II solutions. This is not correct.
Defining type II solutions by 
$f_{10}(r)=\hat{H}_{1,\theta\theta}=0$ on the $z$-axis (for $n=2$)
\cite{han2}, Hannibal then
argues, that $f_{10}(r)$ specifies a boundary condition, which can be chosen 
freely, and that therefore a non-zero $\hat{H}_{1,\theta\theta}$ on the 
$z$-axis can not be generated. However, only for a local solution 
can the function $f_{10}(r)$ be chosen freely, i.~e.~for any function 
$f_{10}(r)$ there exists a solution of the differential equations which is
valid only in a region near the $z$-axis. For a global solution the function 
$f_{10}(r)$ has to be chosen such that the solution fulfills the correct 
boundary condition at the $\rho$-axis. If the global solution is unique, the 
function $f_{10}(r)$ is fixed (provided the gauge is fixed).


\begin{thebibliography}{000}
\bibitem{YMD}
B. Kleihaus and J. Kunz,
Axially symmetric multisphalerons in Yang-Mills-dilaton theory,
Phys.~Lett. {\bf B392} (1997) 135.
\bibitem{EYMD1}
B. Kleihaus and J. Kunz,
Static axially symmetric solutions of Einstein-Yang-Mills-dilaton theory,
Phys.~Rev.~Lett. {\bf 78} (1997) 2527.
\bibitem{EYMD2}
B. Kleihaus and J. Kunz,
Static axially symmetric Einstein-Yang-Mills-dilaton solutions: Regular 
solutions,
Phys.~Rev. D {\bf 57} (1998) 834.
\bibitem{regu}
B. Kleihaus,
The regularity of static axially symmetric solutions in $SU(2)$ 
Yang-Mills-dilaton theory,
Phys.~Rev. D {\bf 59} (1999) 125001. 
\bibitem{EYMDBH1}
B. Kleihaus and J. Kunz,
Static black hole solutions with axial symmetry,
Phys.~Rev.~Lett. {\bf 79} (1997) 1595.
\bibitem{EYMDBH2}
B. Kleihaus and J. Kunz, Static axially symmetric 
Einstein-Yang-Mills-dilaton solutions. II. Black hole solutions,
Phys.~Rev. D {\bf 57} (1998) 6138.
\bibitem{han1}
L. Hannibal,
Singularities in axially symmetric solutions of Einstein-Yang-Mills and
related theories, hep-th/9903063.
\bibitem{RR}
C. Rebbi and P. Rossi,
Multimonopole solutions in the Prasad-Sommerfield limit,
Phys.~Rev. D {\bf 22} (1980) 2010.
\bibitem{han2}
L. Hannibal,
Existence of axially symmetric solutions in SU(2)-Yang Mills and related 
theories,
hep-th/9907222.
\bibitem{rep1}
B. Kleihaus and J. Kunz,
Comment on ``Singularities in axially symmetric solutions of 
Einstein-Yang-Mills and related theories, by Ludger Hannibal, 
[hep-th/9903063]'',
preprint hep-th/9903235. 
\bibitem{others}
G. Lavrelashvili and D. Maison,
Static spherically symmetric solutions of a Yang-Mills field coupled
to a dilaton,
Phys.~Lett. {\bf B295} (1992) 67;\\
P. Bizon,
Saddle-point solutions in Yang-Mills-dilaton theory,
Phys.~Rev. D {\bf 47} (1993) 1656;\\
R. Bartnik and J. McKinnon,
Particle like solutions of the Einstein-Yang-Mills equations,
Phys.~Rev.~Lett. {\bf 61} (1988) 141;\\
N. Straumann and Z.~H. Zhou,
Instability of the Bartnik-McKinnon solutions of the Einstein-Yang-Mills
equations,
Phys.~Lett. {\bf B237} (1990) 353;\\
D.~V. Gal'tsov and M.~S. Volkov,
Sphalerons in Einstein-Yang-Mills theory,
Phys.~Lett. {\bf B273} (1991) 255;\\
M.~S. Volkov and D.~V. Gal'tsov,
Black holes in Einstein Yang-Mills theory,
Sov. J. Nucl. Phys. 51 (1990),747;\\
P. Bizon, Colored black holes,
Phys. Rev. Lett. {\bf 64} (1990) 2844; \\
E.~E. Donets and D.~V. Gal'tsov,
Stringy sphalerons and non-abelian black holes,
Phys.~Lett. {\bf B302} (1993) 411;\\
G. Lavrelashvili and D. Maison,
Regular and black hole solutions of Einstein-Yang-Mills-dilaton theory,
Nucl.~Phys. {\bf B410} 407;\\ 
T. Torii and K. Maeda,
Black holes with non-Abelian hair and their thermodynamical properties,
Phys.~Rev. D {\bf 48} (1993) 1643;\\ 
P. Bizon,
Saddle points of stringy action,
Acta Phys. Pol. B 24 (1993) 1209;\\ 
C.~M. O'Neill,
Einstein-Yang-Mills theory with a massive dilaton and axion: String-inspired
regular and black hole solutions,
Phys.~Rev. D {\bf 50} (1994) 865;\\ 
B. Kleihaus, J. Kunz and A. Sood,
SU(3) Einstein-Yang-Mills-dilaton sphalerons and black holes,
Phys.~Lett. {\bf B372} (1996) 204;\\ 
B. Kleihaus, J. Kunz and A. Sood,
Sequences of Einstein-Yang-Mills-dilaton black holes,
Phys. Rev. D {\bf 54} (1996) 5070. 
\bibitem{proofs}
 J.~A. Smoller and A.~G. Wasserman,
Existence of infinitely many smooth, static, global solutions of the 
Einstein/Yang-Mills equations,  
Commun. Math. Phys. 151 (1993) 303;\\
H.P. Kuenzle,
SU(N) Einstein Yang-Mills fields with spherical symmetry,
Class. Quant. Grav. 8 (1991) 2283.
\bibitem{dil}
We set the dilaton coupling constant $\kappa$ equal to 1.
\bibitem{related}
For pure EYM solutions the constant $\phi_1$ vanishes, whereas for 
YMD solutions the constants 
$\bar{{\rm g}}_1$, $\bar{{\rm l}}_2$ and $\bar{{\rm m}}_2$
vanish.  
\bibitem{asymp} 
B. Kleihaus and Jutta Kunz, in preparation.
The asymptotic solution of \cite{EYMD2} corresponds to
$C_2=0$, whereas the numerical solutions of
\cite{YMD,EYMD1,EYMD2} have $C_2 \ne 0$.
\end{thebibliography}
\end{document}